\documentclass[prl,twocolumn,showpacs,floatfix]{revtex4}
\usepackage{amssymb, amsmath}
\usepackage[dvips]{graphicx}
\usepackage{latexsym}
\usepackage{verbatim}

\begin{document}

\title{Measurement of the current-phase relation of superconducting atomic
contacts}
\author{M.L. Della Rocca, M. Chauvin, B. Huard, H. Pothier, D. Esteve, and
C. Urbina} \affiliation{Quantronics group, Service de Physique de l'\'{E}tat Condens\'{e} (CNRS URA 2464), DSM/DRECAM,
CEA-Saclay, 91191 Gif-sur-Yvette Cedex, France }
\date{\today}

\begin{abstract}
We have probed the current-phase relation of an atomic contact placed with a tunnel junction in a small superconducting
loop. The measurements are in quantitative agreement with the predictions of a resistively shunted SQUID model in which
the Josephson coupling of the contact is calculated using the independently determined transmissions of its conduction
channels.
\end{abstract}

 \maketitle

The Josephson effect is a striking signature of extended quantum coherent states of matter, as found in superfluids,
superconductors, and atomic Bose-Einstein condensates. It appears when a weak link allows particles to flow between two
reservoirs of such quantum systems, thereby establishing phase coherence between the two corresponding macroscopic wave
functions. The effect was first predicted and observed for the case of superconductors, where DC electrical
supercurrents flow in absence of any voltage and AC supercurrents appear under a constant voltage bias
\cite{Josephson}. Since then, it has also been explored in superfluids \cite{Helium} and in Bose-Einstein condensates
\cite{BEC}. Although in the field of superconductivity a large variety of weak links has been used (tunnel junctions,
proximity effect bridges, point contacts, graphene, carbon nanotubes, \ldots )
\cite{Likharev,Golubov,Kasumov,Doh,Graphene}, the basic effect is generic and a unifying picture, able to treat on the
same footing all the different coupling structures, emerged in the framework of mesoscopic superconductivity. Within
this framework, the basic Josephson weak link is a single conduction channel of arbitrary transmission probability
$\tau $ \cite{Landauer} connecting two superconducting electrodes. For a channel shorter than the superconducting
coherence length, the Josephson coupling between both sides is established by a single pair of ``Andreev bound states''
\cite{Furusaki,Beenakker}. These are described as resonant electron-hole quasiparticles states spreading in both
electrodes, with energies $E_{\tau }^{\pm }(\delta ,\tau )=\pm \Delta \left( 1-\tau \sin ^{2}(\delta /2)\right)
^{1/2}$, $\Delta $ being the superconducting gap, and $\delta $ the phase difference between the order parameters on
both sides. These states carry opposite supercurrents $I_{\tau }^{\pm }(\delta )=2\pi \phi _{0}^{-1}\partial E_{\tau
}^{\pm }/\partial \delta $, where $\phi _{0}=h/2e$ is the flux quantum. At zero temperature only the lower state is
occupied, and the current-phase relation \ for the weak link is simply $I_{\tau }^{-}(\delta )$. The critical current
of a channel $I_{\tau }^{0}=\max \left\{ I_{\tau }^{-}(\delta )\right\} $ is the maximum supercurrent that can be
sustained in absence of fluctuations of the phase. Beyond this value, a voltage develops across the system, i.e.
transport becomes dissipative, and is
perfectly understood in terms of multiple Andreev reflection (MAR) processes %
\cite{BTK}, which also depend strongly on $\tau .$ In general, any phase coherent conductor can be described as a
collection of independent
conduction channels, characterized by its set of transmission coefficients $%
\{\tau _{i}\},$ and the global current-phase relation $I_{\{\tau _{i}\}}(\delta )=\sum {I_{\tau _{i}}^{-}(\delta )}$
for an arbitrary weak link is simply the sum of the contributions of its channels. It is then clearly of fundamental
interest to measure the current-phase relation for a single channel of arbitrary transmission. Atomic contacts of
superconducting metals are suitable systems to test these ideas, even if they tend to comprise not just one but a few
channels \cite{ReviewALR}. A measurement of the current-phase relation of atomic size point contacts was first
performed by Koops \textit{et al.} \cite{Koops}, but a quantitative comparison with theory could not be performed
because the $\{\tau _{i}\}$ were not known.
Since then, a reliable method to determine the transmissions was developed %
\cite{Scheer97}, based on the measurement of the dissipative MAR current under a voltage bias. In this Letter we
present measurements of the current-phase relation of well characterized aluminum atomic contacts, and a direct
comparison with theory, with no adjustable parameter.

\begin{figure}[tbph]
\begin{center}
\includegraphics[angle=-90,width=3.4in]{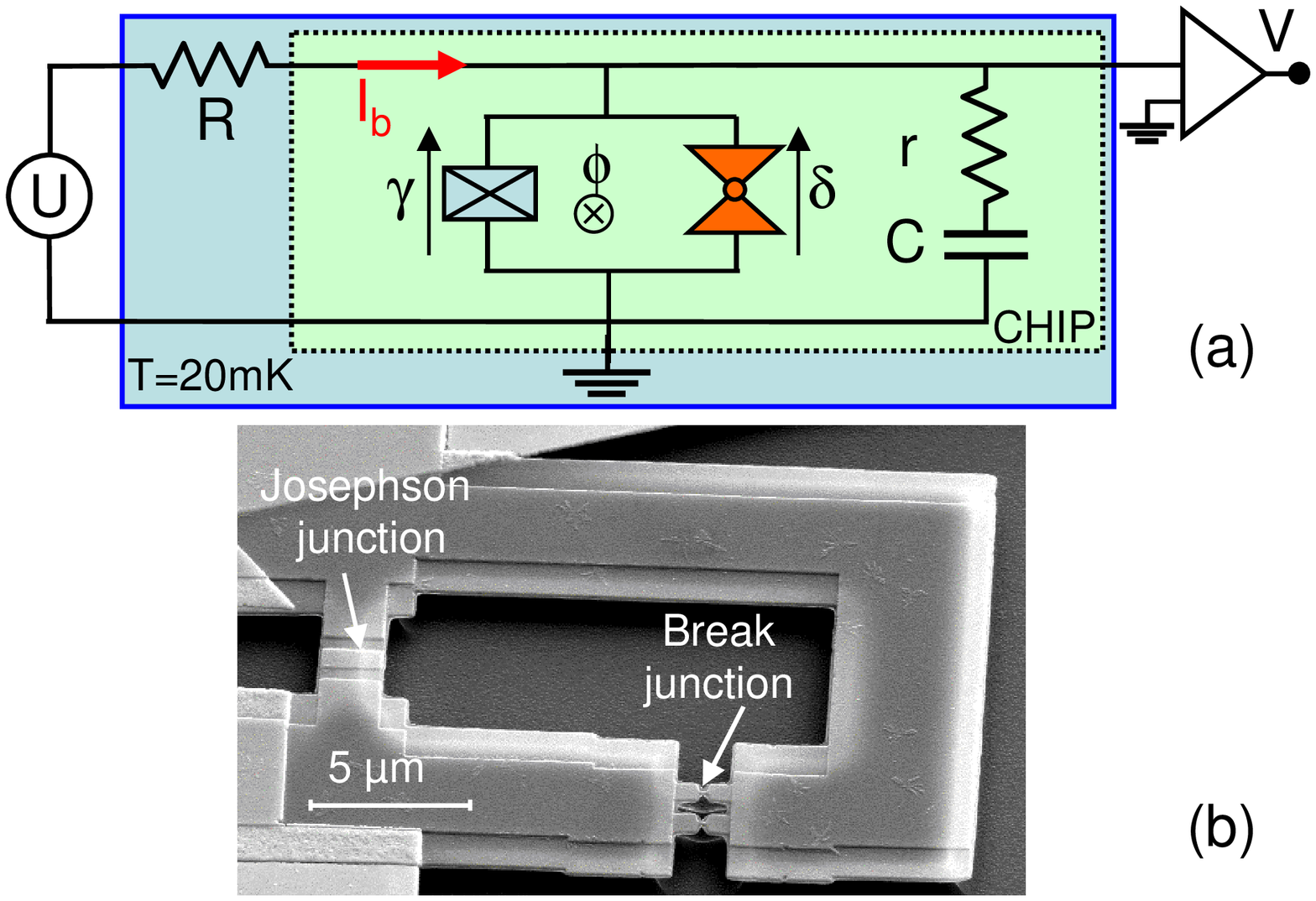}
\end{center}
\caption{(color online) \textbf{(a)} Schematic experimental setup: DC SQUID formed by an atomic contact (phase
$\protect\delta $) and a tunnel junction (phase $\protect\gamma $). The on-chip capacitor $C\approx 21\,\mathrm{pF}$ is
formed between the metallic substrate and a $100\,\mathrm{nm}$-thick gold
electrode ($1.3\,\mathrm{mm}^{2}$), with a $1.6\,\mathrm{\protect\mu m}$%
-thick dielectric polyimide layer. The resistor $r\approx 0.6\,\mathrm{%
\Omega }$, corresponds to the sheet resistance of the capacitor gold electrode. The bias current $I_{b}$ is governed by
voltage source $U$ and discrete macroscopic resistor $R\approx 25\,\mathrm{\Omega }$ mounted close to sample, at base
temperature $\left( 20\,\mathrm{mK}\right) $ of dilution refrigerator. \textbf{(b)} SEM image of SQUID loop. Tunnel
junction is fabricated using double-angle evaporation of aluminum through a suspended mask. This results in a parasitic
structure of no practical importance. Bright regions on left corners correspond to the gold thin films that connect the
superconducting loop to the rest of the circuit, and provide the top plate of the capacitor. These normal regions also
act as quasiparticle traps.} \label{circuit}
\end{figure}

The principle of our experimental setup is shown schematically in Fig.\thinspace 1a, and is designed to allow for both
voltage and phase bias of the samples, giving this way independent access to the transmission probabilities and to the
current-phase relation respectively. The sample consists of an atomic contact (phase difference $\delta $) and a tunnel
junction (phase difference $\gamma $) embedded in parallel in a small superconducting loop \cite{TheseChauvin}, hence
forming an asymmetric SQUID, as shown in the scanning electron microscope image of Fig.\thinspace 1b. Note that similar
\cite{Miyazaki} or related \cite{Marchenkov} setups are being used in other laboratories. The atomic contact is
obtained using a microfabricated break junction \cite{JanQuantro}, and is characterized by a
critical current $I_{\{\tau _{i}\}}^{0},$ typically a few tens of $\mathrm{%
nA,}$ much lower than the critical current of the tunnel junction $%
I_{0}\approx 740\,\mathrm{nA}$. The SQUID is placed in parallel with an on-chip $rC$ circuit, which dominates the
parallel impedance of the line and controls the phase dynamics. The sample, which is fabricated on a metallic substrate
coated with a polyimide layer, is thermally anchored to the mixing chamber of a dilution refrigerator. The bias current
$I_{b}$ is provided by
a room temperature voltage source connected through a series of $50\,\mathrm{%
\Omega }$ attenuators placed at low temperatures, and a discrete macroscopic resistor $R$ mounted at the same
temperature as the sample. Voltage and current in the SQUID are both measured with low-noise voltage amplifiers, the
latter from the voltage drop across the resistor $R$.

The idea behind this setup is twofold \cite{TheseChauvin}. On the one hand, it allows to obtain the dissipative part of
the current-voltage
characteristic of the contact $I_{\{\tau _{i}\}}(V)=I_{\mathrm{SQUID}%
}(V)-I_{J}(V),$ as the difference between the one of the SQUID $I_{\mathrm{%
SQUID}}(V)$ and the one of the tunnel junction $I_{J}(V)$ alone, as shown in the lower panel of Fig.\thinspace 2. The
latter is measured after fully opening the break junction \cite{subgap}. One then determines the transmission
probabilities $\{\tau _{i}\}$ and the gap $\Delta \simeq 180\,\mathrm{\mu eV}
$, by fitting $I_{\{\tau _{i}\}}(V)$ with MAR theory \cite%
{AverinBardas,Bratus95,Cuevas96,Scheer97}, as shown in the upper panel of Fig.\thinspace 2 for three different
contacts. On the other hand, on the supercurrent branch of the SQUID (see inset in Fig.\thinspace 2), it is possible to
impose a phase difference on the contact using both the external flux and the current bias as control knobs, and to use
the tunnel junction as a threshold detector to measure the current flowing in
the loop \cite{quantronium}. Indeed, the loop is designed to be small enough %
\cite{Loop} so that, to a very good approximation, the two phases are linked by the magnetic flux $\phi $ threading the
loop, according to $\delta -\gamma =\varphi =2\pi \phi /\phi _{0}$ and therefore $I_{b}=I_{0}\sin \gamma +I_{\{\tau
_{i}\}}(\gamma +\varphi ).$ In the limit $I_{0}\gg I_{\{\tau _{i}\}}^{0}$ the critical current $I_{c}$ of the SQUID
should be reached when $\gamma \sim \pi /2,$ and therefore its variations with the external flux are $I_{c}\left(
\varphi \right) \sim I_{0}+I_{\{\tau
_{i}\}}(\varphi +\pi /2).$ Therefore, the periodic modulations of $%
I_{c}\left( \varphi \right) $ around the critical current of the junction probe directly the current-phase relation of
the atomic contact. It is however important to note that in practice, due to fluctuations, both quantum and thermal,
the system ``switches'' stochasticaly from the supercurrent branch to the dissipative branch before the bias current
reaches $I_{c}.$ This switching process is characterized by a rate $\Gamma .$ It is nevertheless still possible, as we
show hereafter, to probe the current-phase relation of the contact from measurements of the switching current of the
whole device as a function of the magnetic flux.

\begin{figure}[tbph]
\begin{center}
\includegraphics[width=3.4in]{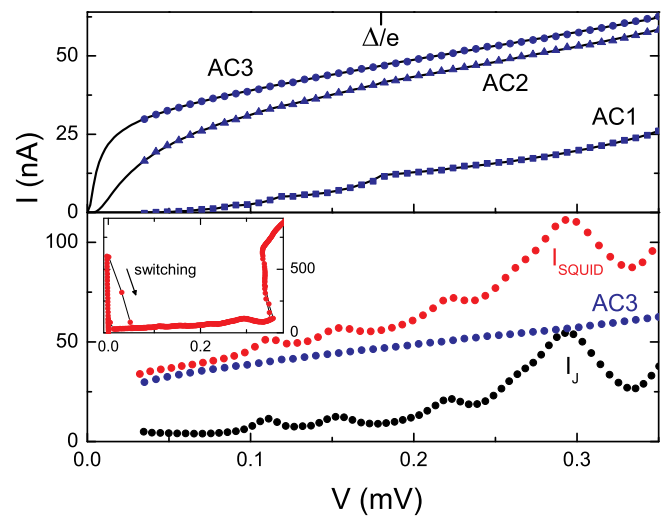}
\end{center}
\caption{(color online) \textbf{Upper panel:} Experimental current-voltage curves for three atomic contacts (symbols)
for $0<eV<2\Delta ,$ compared to best fits $I_{\{\protect\tau _{i}\}}(V)$ using MAR (full lines). Fits
provide gap $\Delta \simeq 180\,\mathrm{\protect\mu eV}$ and set $\{\protect%
\tau _{i}\}$ of transmission coefficients for each contact: $\mathrm{AC1}%
\rightarrow \{0.62;0.22;0.07\}$; $\mathrm{AC2}\rightarrow \{0.957;0.19\}$; $%
\mathrm{AC3}\rightarrow \{0.993;0.14\}$. \textbf{Lower panel:} $I(V)$ curve
for contact AC3, obtained as the difference between $I_{SQUID}(V)$ and $%
I_{J}(V)$. \textbf{%
Inset: }Large scale characteristic of SQUID, displaying supercurrent branch at $V=0$. } \label{IV}
\end{figure}

\begin{figure}[tbph]
\begin{center}
\includegraphics[width=3.4in]{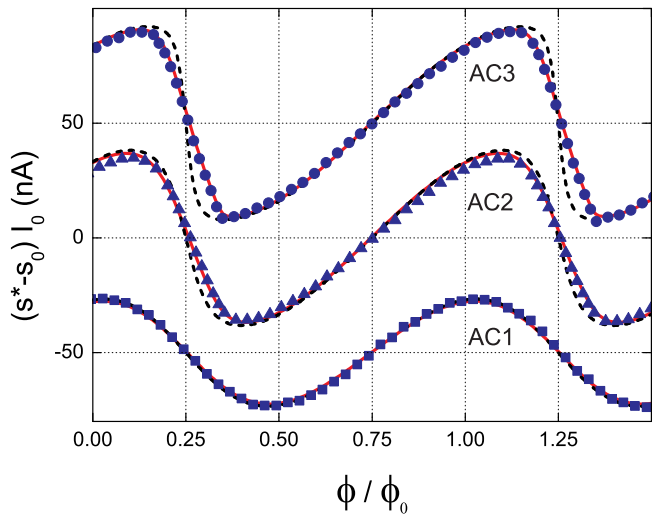}
\end{center}
\caption{(color online) \textbf{Symbols}: measured switching current $%
(s^{\ast }(\protect\varphi )-s_{0})I_{0}$ as a function of applied flux $%
\protect\phi /\protect\phi _{0},$ for the three SQUIDS corresponding to the contacts of Fig. 2. Curves AC3 and AC1
shifted for clarity. \textbf{Dashed
curves: } predicted ground state current-phase relation $I_{\{\protect\tau %
_{i}\}}^{-}(\protect\delta )$. \textbf{Full lines:} predictions of resistively shunted SQUID theory at
$T_{\mathrm{esc}}=130\,\mathrm{mK}$ on the basis of Eq.~(1). The transmission sets indicated in Fig.~2 caption have
been used for both theories.} \label{Iphase}
\end{figure}

Typically, we apply $10^{4}$ bias current pulses of amplitude $s=I_{b}/I_{0}$ and duration $\tau _{p}\sim
40\,\mathrm{\mu s}$, and measure the switching probability $P(s)=1-e^{-\Gamma (s)\tau _{p}}$ as the ratio between the
number
of switching events and the total number of pulses. For each value of $%
\varphi $ we adjust the current pulse amplitude $s^{\ast }(\varphi )$ so as to keep a constant switching probability
$P(s^{\ast })=0.6$ (corresponding to a rate $\Gamma ^{\ast }=23.3\,\mathrm{kHz}$), which leads to the best sensitivity
with respect to flux response. The $s^{\ast }(\varphi )$ curves measured in this way for the three contacts of
Fig.\thinspace 2 are shown as symbols in Fig.\thinspace 3. As the absolute value of the flux through the loop is not
known, and the current measurements suffer from small drifting offsets, the experimental curves had to be shifted in
both directions by an arbitrary amount in order to compare them with the theoretical predictions.
Note however that the vertical shift corresponds essentially to the average value $%
\left\langle s^{\ast }(\varphi )\right\rangle ,$ which is very close to the switching current $s_{0}$ of the junction
alone under the same conditions.
Figure\thinspace 3 also shows the calculated current-phase relations $%
I_{\{\tau _{i}\}}(\varphi +\pi /2)$ for the corresponding sets $\{\tau _{i}\}$. There is an overall qualitative
agreement between the experimental data and these simple predictions. The discrepancies are significant only for
contacts AC2 and AC3, which both contain a highly transmitted channel, and arise mainly around a phase $\delta =\pi .$
These differences can be understood almost completely by taking into account the phase fluctuations imposed by the
dissipative elements of the electromagnetic environment in which the SQUID is embedded. It is well known that in such a
dissipative biasing circuit the phase across the SQUID is a dynamical variable governed by a Langevin equation,
equivalent to the one obeyed by the position of a massive particle evolving in a ``tilted washboard potential'' in
presence of friction \cite{Barone}. Assuming that only the ground Andreev state of each channel of the atomic contact
is occupied, the total potential of the SQUID
is given by:%
\begin{equation}
U_{-}(\gamma )=-E_{J}\cos \gamma -E_{J}s\gamma -\sum_{i}E_{\tau _{i}}^{-}(\gamma +\varphi ,\tau _{i})
\end{equation}%
where the first term is the Josephson energy of the tunnel junction, with $%
E_{J}=\phi _{0}I_{0}/2\pi $, the second one is the energy arising from the coupling to the current source, and the last
term is the Josephson coupling introduced by the atomic contact. Figure\thinspace 4 shows $U_{-}(\gamma )$ for a SQUID
with a single channel contact $(\tau =0.99)$, and for comparison the potential of the tunnel junction alone.

\begin{figure}[tbph]
\begin{center}
\includegraphics[width=3.4in]{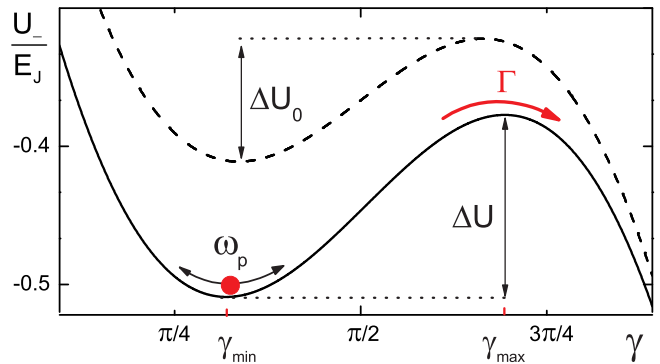}
\end{center}
\caption{(color online) \textbf{Full line:} washboard potential of a SQUID
with a single channel contact for $\protect\tau =0.99,$ $s=0.87$ and $%
\protect\varphi =0$, as function of the Josephson junction phase $\protect%
\gamma $. Thermal activation allows the phase to escape at a rate $\Gamma $ above the barrier of height $\Delta U$.\
\textbf{Dashed line:} washboard potential of Josephson tunnel junction alone for the same parameters.}
\label{Potential}
\end{figure}

The overall shape of the potentials is qualitatively the same but for very highly transmitted channels ($\tau >0.999$),
and the physics is therefore similar to the well known case of tunnel junctions. For the actual
parameters of the setup, one can neglect quantum fluctuations and treat $%
\gamma $ as a classical variable. For $0<s<1$, the equivalent particle
oscillates around a local minimum of the potential at the plasma frequency $%
\omega _{p}(s)=\omega _{0}(1-s^{2})^{1/4}$, with $\omega _{0}\simeq (2\pi I_{0}/\phi _{0}C)^{1/2}$ \cite{plasma}. The
tilt of the potential increases with $s$, and the thermal energy $k_{B}T$ becomes eventually comparable to
the potential barrier height $\Delta U(\gamma )={U_{-}(\gamma _{\mathrm{max}%
})-U_{-}(\gamma _{\mathrm{min}})}$, where $\gamma _{\mathrm{min}}$ ($\gamma _{\mathrm{max}}$) is the phase at which the
potential presents a local minimum (maximum). The particle can then be thermally activated over the barrier and escape
from the well at a rate
\begin{equation}
\Gamma (s,\varphi )\simeq \frac{\omega _{p}(s)}{2\pi }\ e^{-\Delta U(s,\varphi )/k_{B}T},
\end{equation}%
before the current actually reaches the critical current. The biasing circuit is such that, once escaped, the particle
runs away indefinitely and a voltage suddenly develops at the edge of the SQUID, according to $V=\phi
_{0}\dot{\gamma}/2\pi .$ This corresponds to the ``switching'' detected in the experiments.

We first performed switching measurements of the Josephson junction alone, which is a well known case \cite{Barone}.
The reduced bias current corresponding to the imposed escape rate
is of the order of $s_{0}=0.87$, corresponding to a phase ${\gamma _{\mathrm{%
min}}}=\gamma _{0}=\mathrm{arc}\sin \left( s_{0}\right) \simeq 0.67\left( \pi /2\right) $.
The s-dependence of the switching rate agrees precisely with Eq. (2), and yields the escape temperature $T_{\mathrm{esc}%
}\approx 125\,\mathrm{mK}$. Although it is significantly higher than the refrigerator temperature
$T_{0}=20\,\mathrm{mK},$ showing that the electrons in the dissipative elements of the biasing circuit are heated by
some remaining spurious noise, it does not hinder the measurement of the current-phase relation. To compare the
experimental data for the SQUIDs with theory, we assume this measured $T_{\mathrm{esc}}$ to be the actual temperature
determining the phase fluctuations in all cases, and given the exponential dependence of the rate on $\Delta
U(s,\varphi )$, that a constant escape rate corresponds to a constant barrier height. We then use
Eq.\thinspace (1), with the $\{\tau _{i}\}$ obtained independently (Fig.\ref%
{IV}), to calculate, for each SQUID, the current $s^{\ast }(\varphi )$ leading to the imposed rate $\Gamma ^{\ast }$.
The curves calculated in this manner are shown as full lines in Fig.\thinspace 3 and describe the experimental data
significantly better than the $T=0$ theory. Note that this procedure assumes that in each channel only the ground
Andreev state is occupied, and that the only effect of the finite temperature is on the dynamics of the phase. This is
not forcedly the case for the contacts measured here, as $k_{B}T_{\mathrm{esc}}$ is not much smaller than the minimum
energy gap $2\Delta \left( 1-\tau \right) ^{1/2}$ between Andreev levels at $\delta =\pi $, but the resulting
population of the excited state is still too small to have a detectable consequence.

In conclusion, we have measured the current-phase relation of superconducting atomic contacts covering a wide range of
transmission coefficients, and accounted quantitatively for the results using the mesoscopic theory of the Josephson
effect. Note that the experiments described here probe just the ground Andreev state of each channel but that it should
be possible and interesting to also probe the excited state, through microwave spectroscopy for instance. Furthermore,
by designing a proper environment such that the life-time of this excited state and the dephasing time between the two
states are long enough, one could envision to create coherent quantum superpositions of them\thinspace
\cite{Lantz2002}.

\begin{acknowledgments}
This work was supported by the EU-RTN DIENOW. We thank important discussions with J. Clarke, A. Levy Yeyati, V.
Shumeiko and C. Strunk. Constant help from the other members of the Quantronics group, specially Q. Le Masne, is
gratefully acknowledged.
\end{acknowledgments}


\begin{thebibliography}{99}
\bibitem{subgap}The finite and peaked sub-gap current is attributed to
inelastic Cooper pair transfer with excitation of electromagnetic resonances in the environment \cite{Holst}. This
structure was independent of the bending of the substrate. Note that large sub-gap currents have been also observed in
similar junctions fabricated on silicon \cite{Wellstood,Lupascu}.
\bibitem{Loop} The geometrical inductance of the loop $L\approx 10\,\mathrm{%
pH}$ is negligible compared to the Josephson inductances of the junction $%
L_{J}=\phi _{0}/2\pi I_{0}\approx 1\,\mathrm{nH}$ and of the atomic contact $%
L_{AC}\approx 10\,\mathrm{nH}.$
\bibitem{plasma} The plasma frequency of the junction $\left( \omega
_{0}/2\pi \simeq 1.62~\mathrm{GHz}\right) $ and the losses measured independently through microwave reflectometry, are
in good agreement with the measured value of $I_{0}$, and the values of $C$ and $r$ estimated from the geometry. With
these parameters, the dynamics of the phase around the plasma frequency is underdamped.
\end{thebibliography}
\end{document}